\documentclass[useAMS,usenatbib]{mn2e}
\usepackage{graphicx}

\newcommand{\Teff}{\mbox{$T_{\mathrm{eff}}$}}
\newcommand{\Line}[3]{\Ion{#1}{#2}~#3\,\AA}
\newcommand{\Lines}[3]{\Ion{#1}{#2}\,#3\,\AA}
\newcommand{\Ion}[2]{#1{\,\sc#2}}
\newcommand{\La}{\mbox{${\mathrm{Ly\alpha}}$}}
\newcommand{\kms}{\mbox{$\mathrm{km\,s^{-1}}$}}

\newcommand{\Rsun}{\mbox{$\mathrm{R}_{\odot}$}}

\title{The chemical diversity of exo-terrestrial planetary debris around white
  dwarfs}

\author[B.T. G\"ansicke et al.]{
B.T. G\"ansicke$^1$,
D. Koester$^2$,
J. Farihi$^3$,
J. Girven$^1$,
S.G. Parsons$^1$,
E. Breedt$^1$\\
$^{1}$ Department of Physics, University of Warwick, Coventry CV4 7AL,
UK \\
$^{2}$ Institut f\"ur Theoretische Physik und Astrophysik, University of Kiel,
24098 Kiel, Germany\\
$^{3}$ Department of Physics Astronomy, University of Leicester,
Leicester LE1 7RH, UK
}

\begin{document}

\date{Accepted 2005. Received 2005; in original form 2005}

\pagerange{\pageref{firstpage}--\pageref{lastpage}} \pubyear{2006}

\maketitle

\label{firstpage}

\begin{abstract}
We present \textit{Hubble Space Telescope} ultraviolet spectroscopy of
the white dwarfs PG\,0843+516, PG\,1015+161, SDSS\,1228+1040, and
GALEX\,1931+0117, which accrete circumstellar planetary debris formed
from the destruction of asteroids. Combined with optical data, a
minimum of five and a maximum of eleven different metals are detected
in their photospheres. With metal sinking time scales of only a few
days, these stars are in accretion/diffusion equilibrium, and the
photospheric abundances closely reflect those of the circumstellar
material. We find C/Si ratios that are consistent with that of the
bulk Earth, corroborating the rocky nature of the debris. Their C/O
values are also very similar to those of bulk Earth, implying that the
planetary debris is dominated by Mg and Fe silicates. The abundances
found for the debris at the four white dwarfs show substantial
diversity, comparable at least to that seen across different meteorite
classes in the solar system. PG\,0843+516 exhibits significant
over-abundances of Fe and Ni, as well as of S and Cr, which suggests
the accretion of material that has undergone melting, and possibly
differentiation. PG\,1015+161 stands out by having the lowest Si
abundance relative to all other detected elements.  The Al/Ca ratio
determined for the planetary debris around different white dwarfs is
remarkably similar. This is analogous to the nearly constant abundance
ratio of these two refractory lithophile elements found among most
bodies in the solar system.

Based on the detection of all major elements of the circumstellar
debris, we calculate accretion rates of
$\simeq1.7\times10^8\,\mathrm{g\,s^{-1}}$ to
$\simeq1.5\times10^9\,\mathrm{g\,s^{-1}}$. Finally, we detect
additional circumstellar absorption in the \Lines{Si}{iv}{1394,1403}
doublet in PG\,0843+516 and SDSS\,1228+1040, reminiscent to similar
high-ionisation lines seen in the \textit{HST} spectra of white dwarfs
in cataclysmic variables. We suspect that these lines originate in hot
gas close to the white dwarf, well within the sublimation radius.
\end{abstract}


\section{Introduction}
Most of our current insight into the interior structure of exo-planets
is derived from the bulk density of transiting planets
\citep[e.g.][]{valenciaetal10-1}, and transit spectroscopy provides
some information on the chemical composition of their atmospheres
\citep[e.g.][]{grillmairetal08-1}.  More detailed investigations of
the chemistry of exo-planetary systems around main-sequence host stars
are beyond the reach of present observational
instrumentation. However, \citet{zuckermanetal07-1} demonstrated in a
pioneering paper that the photospheric abundances of polluted white
dwarfs can be used to infer the bulk abundances of the planetary
debris material detected around the white dwarf GD\,362, and showed
that the composition of this material is broadly comparable to that of
the Earth-Moon system.

The strong surface gravity of white dwarfs implies that metals will
sink out of the photosphere on time scales that are orders of
magnitude shorter than their cooling ages, and therefore white dwarfs
are expected to have either pure hydrogen or helium atmospheres
\citep{fontaine+michaud79-1}. Exceptions to this rule are only hot
($\Teff\ga25\,000$\,K) white dwarfs where radiative levitation can
support some heavy elements in the photosphere
\citep[e.g.][]{chayeretal95-1}, and cool ($\Teff\la10\,000$\,K) white
dwarfs where convection may dredge up core material
\citep{koesteretal82-2, fontaineetal84-1}. Yet white dwarfs with
metal-contaminated atmospheres have been known for nearly a century
\citep{vanmaanen17-1}, and accretion from the interstellar medium
\citep[e.g.][]{koester76-1, wesemael79-1, dupuisetal93-2} has been the
most widely accepted scenario, despite a number of fundamental
problems \citep[e.g.][]{aannestadetal93-1, friedrichetal04-1,
  farihietal10-2}. However, the rapidly growing number of white dwarfs
that are accreting from circumstellar discs
\citep[e.g.][]{becklinetal05-1, kilicetal05-1, gaensickeetal06-3,
  vonhippeletal07-1, farihietal08-1, vennesetal10-1, dufouretal12-1}
unambiguously demonstrates that debris from the tidal disruption of
main-belt analogue asteroids or minor planets \citep{grahametal90-1, jura03-1},
or Kuiper-belt like objects \citep{bonsoretal11-1}, likely perturbed by
unseen planets \citep{debesetal02-1, debesetal12-1}, is the most
likely origin of photospheric metals in many, if not most polluted
white dwarfs.

Because of the need for high-resolution, high-quality spectroscopy,
detailed abundance studies have so far been limited to a handful of
white dwarfs \citep{kleinetal10-1, kleinetal11-1, vennesetal11-1,
  melisetal11-1, zuckermanetal11-1, dufouretal12-1, juraetal12-1}. For
a given abundance and white dwarf temperature, metal lines are
stronger in a helium-dominated (DB) atmosphere than in a
hydrogen-dominated (DA) atmosphere, as the opacity of helium is much
lower than that of hydrogen. Therefore, the small sample of
well-studied metal polluted white dwarfs is heavily biased towards DB
white dwarfs, which have diffusion time scales of
$\sim10^5-10^6$\,yr. These long diffusion time scales introduce a
significant caveat in the interpretation, as the abundances of the
circumstellar debris may substantially differ from those in the white
dwarf photosphere if the accretion rate varies on shorter time scales
\citep{koester09-1}. While the life times of the debris discs are
subject to large uncertainties, there are theoretical
\citep{rafikov11-2, metzgeretal12-1} and observational
(\citealt{girvenetal12-1}, Farihi et al. 2012 in press) arguments
that suggest that the accretion rates onto the white dwarfs may vary
significantly over periods that are short compared to the diffusion
time scales. In fact, some of the most heavily polluted white dwarfs
have no infrared excess \citep{farihietal09-1, kleinetal11-1}, and may
have accreted all the circumstellar debris a few diffusion time scales
ago \citep{farihietal09-1, girvenetal12-1}.

We are currently carrying out an ultraviolet spectroscopic survey of
young DA white dwarfs that have cooling ages of 20 to 200\,Myr, metal
sinking time scales of a few days, and are hence guaranteed to be in
accretion-diffusion equilibrium. The aim of this survey is to
determine the fraction of white dwarfs that are presently accreting
planetary debris, and to determine accurate abundances for a
subset. Here we present the analysis of four heavily polluted white
dwarfs that are known to also host planetary debris discs.

\begin{figure*}
\centerline{\includegraphics[angle=270,width=2\columnwidth]{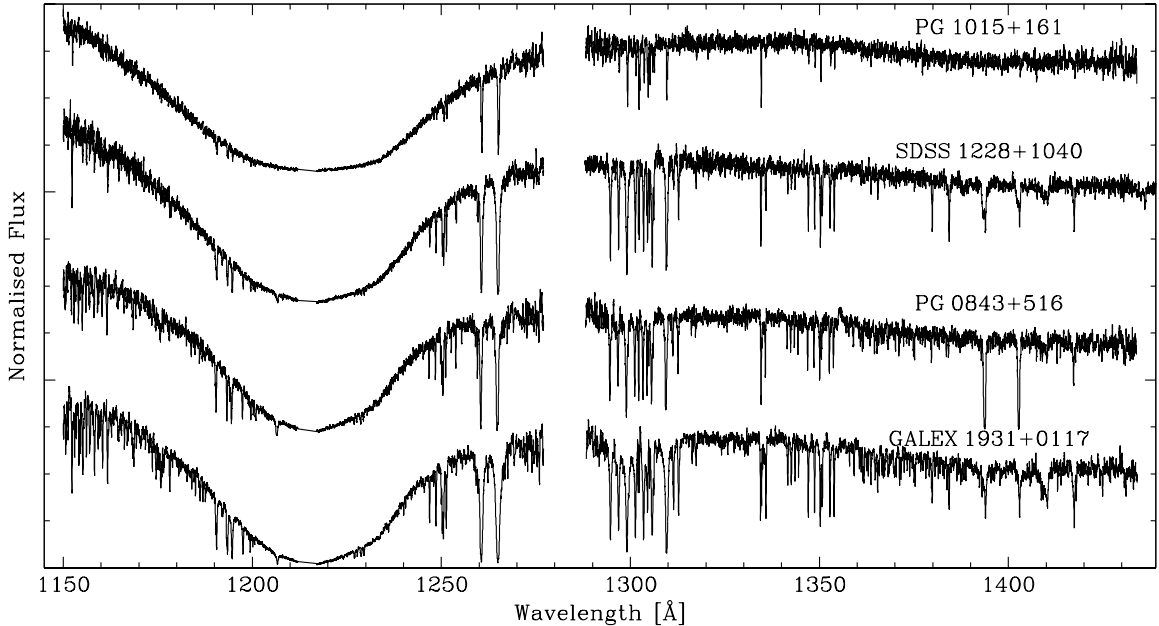}}
\caption{\label{f-cos} COS/G130M spectra of four white dwarfs known to
  have circumstellar discs, scaled to a peak flux of unity, offset by
  1.4 units, and sorted from top to bottom by increasing metal
  abundances. For warm white dwarfs with pure hydrogen atmospheres,
  the broad \La\ line is the only spectral feature in this wavelength
  range. These four stars accrete from the circumstellar debris, and
  their spectra are riddled with absorption lines of C, O, Al, Si, P,
  S, Cr, Fe, and Ni. In addition, Mg and Ca can be detected in their
  optical spectra (Fig.\,\ref{f-camg}).}
\end{figure*}

\section{Observations}

The targets for our ongoing far-ultraviolet spectroscopic survey of
young and correspondingly warm
($17\,000\,\mathrm{K}<\Teff<25\,000$\,K) DA white dwarfs were drawn
from the compilations of \citet{liebertetal05-1} and
\citet{koesteretal09-2}, supplemented with a few recent discoveries
\citep[e.g.][]{gaensickeetal06-3, vennesetal10-1}. Our sample also
includes a small number of post-common envelope binaries (PCEBs) in
which the white dwarf accretes from the wind of the M-dwarf
companion. These systems were selected from
\citet{schreiber+gaensicke03-1} and \citet{farihietal10-3} with the
same cut on white dwarf temperature and cooling age.  Under the
assumption that the M-dwarfs have a solar-like composition, the white
dwarfs in PCEBs serve as ``abundance standards'' for our abundances
analyses and diffusion calculations.

\subsection{\textit{HST}/COS spectroscopy}
\label{s-hstobs}

PG\,0843+516, PG\,1015+161, and GALEX\,J193156.8+011745 (henceforth
GALEX\,1931+0117) were observed as part of our snapshot survey, with
exposure times of 1420\,s, 1424\,s, and 800\,s, respectively.
We used the G130M grating with a central wavelength of 1291\,\AA,
which covers the wavelength range $1130-1435$\,\AA, with a gap at
$1278-1288$\,\AA\ due to the space between the two detector
segments. To mitigate the fixed pattern noise that is affecting the
COS far-ultraviolet detector, we split the exposure time equally
between two FP-POS positions (1\,\&\,4, the limited duration of the
snapshot visits did not allow to use the full set of four different
FP-POS positions).

We also report COS observations of three PCEBs observed within this
snapshot survey, that will be used as ``abundances standards'':
GD\,448 (HR\,Cam, \citealt{maxtedetal98-1}), GD\,245 (MS\,Peg, 
\citealt{schmidtetal95-3}), and PG\,2257+162 (KUV\,22573+1613,
\citealt{wachteretal03-1}), with exposure times of 900\,s, 600\,s,
and 1070\,s, respectively.

SDSS\,J122859.93+104032.9 (henceforth SDSS\,1228+1040) was observed in
Cycle\,17 as part of a regular Guest Observer programme. We obtained
two sets of spectroscopy with the G130M grating with central
wavelengths of 1291\,\AA\ and 1327\,\AA, and both observations were
again split among two FP-POS positions (1\,\&\,4). In addition, we
obtained G160M spectroscopy with central wavelengths of 1577\,\AA\ and
1623\,\AA. The total exposure time of the G130M and G160M observations
were 2821\,s and 4899\,s, respectively, seamlessly covering the
wavelength range $1130-1795$\,\AA.

The data retrieved from the \textit{HST} archive were processed and
calibrated with CALCOS 2.15.6. The COS spectra of the four white
dwarfs shown in Fig.\,\ref{f-cos} reveal the broad \La\ profile
typical of DA white dwarfs, plus a multitude of narrow absorption
lines from a range of metals. The peak signal-to-noise ratio (S/N) of
the COS spectra is reached in a line-free region near $1320$\,\AA, and
ranges from $\simeq25$ for PG\,0843+516 and PG\,1015+161 to $\simeq40$
for SDSS\,1228+1040 and GALEX\,1931+0117. However, these values only
include photon count statistics, and do not account for the residual
fixed-pattern noise related to the use of only two FP-POS
positions. The resolving power of the COS spectra, as measured from
on-orbit data ranges from $\sim15\,000$ at 1150\,\AA\ to $\sim20\,000$
at 1430\,\AA.

\subsection{Optical observations}
\label{s-optobs}
The wavelength spanned by our COS observations does not cover any
strong line of either Ca (traditionally the most important tracer of
metal pollution in white dwarfs, and an important refractory element)
or Mg (one of the major constituents of rocky material in the solar
system, including the Earth). Ground-based abundance studies using the
\Ion{Ca}{ii}~H/K doublet and the \Line{Mg}{ii}{4482} line are already
published for GALEX\,1931+0117 \citep{vennesetal10-1, vennesetal11-1,
  melisetal11-1}. Two short (10\,min) VLT/UVES spectra of PG\,1015+161
were obtained as part of the SPY project \citep{napiwotzkietal01-1},
which \citet{koesteretal05-2} analysed to determine the Ca abundance
of PG\,1015+161 (Sect.\,\ref{s-pg1015}). Here we use the same spectra
to determine in addition the abundance of Mg.

We observed PG\,0843+516 for a total of 2\,h on the WHT using ISIS
with the R600B grating and a $1\arcsec$ slit, covering the Ca and Mg
lines at a resolving power of $\simeq2500$ and a S/N of
$\approx90$. The data were reduced and calibrated as described in
\citet{pyrzasetal12-1}.

We also obtained a total of 9\,h VLT/UVES spectroscopy of
SDSS\,1228+1040 between 2007 and 2009 using the Blue390 and Blue437
setup with a $0.9\arcsec$ slit, covering both the Ca and Mg features
with a resolving power of $\simeq40\,000$. The data were reduced in
Gasgano using the UVES pipeline. The individual spectra were of
relatively low S/N, and we analysed only the error-weighted average
spectrum, binned to 0.05\,\AA, with $\mathrm{S/N}\simeq35$.

\begin{figure}
\includegraphics[angle=-90,width=\columnwidth]{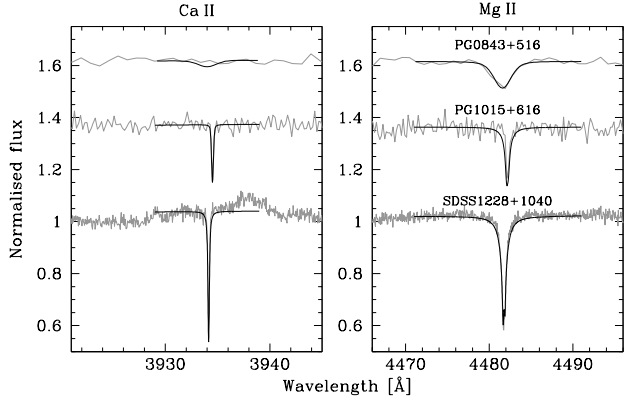}
\caption{\label{f-camg} The normalised optical spectra (gray) of
  PG\,0843+516 (WHT), PG\,1015+161 (VLT/UVES, binned to 0.2\,\AA), and
  SDSS\,1228+1040 (VLT/UVES, binned to 0.05\,\AA), along with the
  best-fit models (black). The width of the \Ion{Mg}{ii} line in
  PG\,0843+516 is due to the low resolution of the WHT data. In the
  spectrum of SDSS\,1228+1040, the photospheric absorption of
  \Ion{Ca}{ii}\,K is embedded in a double-peaked emission line from
  the gaseous debris disc, which is, however, so broad that it does
  not affect the measurement of the Ca abundance.}
\end{figure}

The optical spectra around the \Ion{Ca}{ii}\,K and \Line{Mg}{ii}{4482}
lines are shown in Fig.\,\ref{f-camg}. We note that while most
previous studies of metal-polluted white dwarfs have focused on the
\Ion{Ca}{ii}~H/K lines, their strength for a given abundance decreases strongly
with increasing temperature, as \Ion{Ca}{ii} is ionised to
\Ion{Ca}{iii}. For temperatures $\Teff\simeq20\,000-25\,000$\,K,
\Line{Mg}{ii}{4482} becomes a more sensitive probe of metal pollution
(e.g. \citealt{gaensickeetal07-1, farihietal12-1}).

\begin{table}
 \centering
 \caption{Atmospheric parameters from spectroscopy \label{t-parameters}}
 \begin{tabular}{lrr}
 \hline
  Object & \Teff\ [K]  & $\log g$ [cgs units]     \\
 \hline
\multicolumn{3}{l}{\textbf{PG\,0843+516\,=\,WD\,0843+516}}\\
optical, \cite{liebertetal05-1}& $23\,870 \pm 392$& $7.90 \pm 0.05$\\
HST, this paper                & $23\,095 \pm 230$& $8.17 \pm 0.06$ \\
\multicolumn{3}{l}{\textbf{PG\,1015+161\,=\,WD\,1015+161}}\\
optical, \cite{liebertetal05-1}&$19\,540 \pm 305$ & $8.04 \pm 0.05$\\
optical, \cite{koesteretal09-2}&$19\,948 \pm  33$ & $7.925\pm0.006$\\
HST, this paper                      &$19\,200 \pm 180$& $8.22 \pm 0.06$\\ 
\multicolumn{3}{l}{\textbf{SDSS\,J122859.93+104032.9\,=\,WD1226+110}}\\
optical, \cite{eisensteinetal06-1}   &$22\,125 \pm 136$& $8.22 \pm 0.02$\\
optical, \cite{gaensickeetal07-1}    &$22\,292 \pm 296$& $8.29 \pm 0.05$\\
optical, our fit to SDSS spectrum    &$22\,410 \pm 175$& $8.12 \pm 0.02$\\
HST, this paper                      &$20\,565 \pm 82$ & $8.19 \pm 0.03$\\
adopted, this paper (Sect.~\ref{s-teff_logg}) &$20900 \pm 900$& $8.15 \pm 0.04$\\
\multicolumn{3}{l}{\textbf{GALEX\,J193156.8+011745\,=\,WD\,1929+012}}\\
optical, \cite{vennesetal10-1}   & $20\,890 \pm 120$& $7.90 \pm 0.03$ \\
optical, \cite{melisetal11-1}    & $23\,470 \pm 300$& $7.99 \pm 0.05$ \\
HST, this paper                  & $21\,200 \pm 50$ & $7.91 \pm 0.02$  \\
\hline
\end{tabular}
\end{table}

\begin{figure*}
\includegraphics[width=\columnwidth]{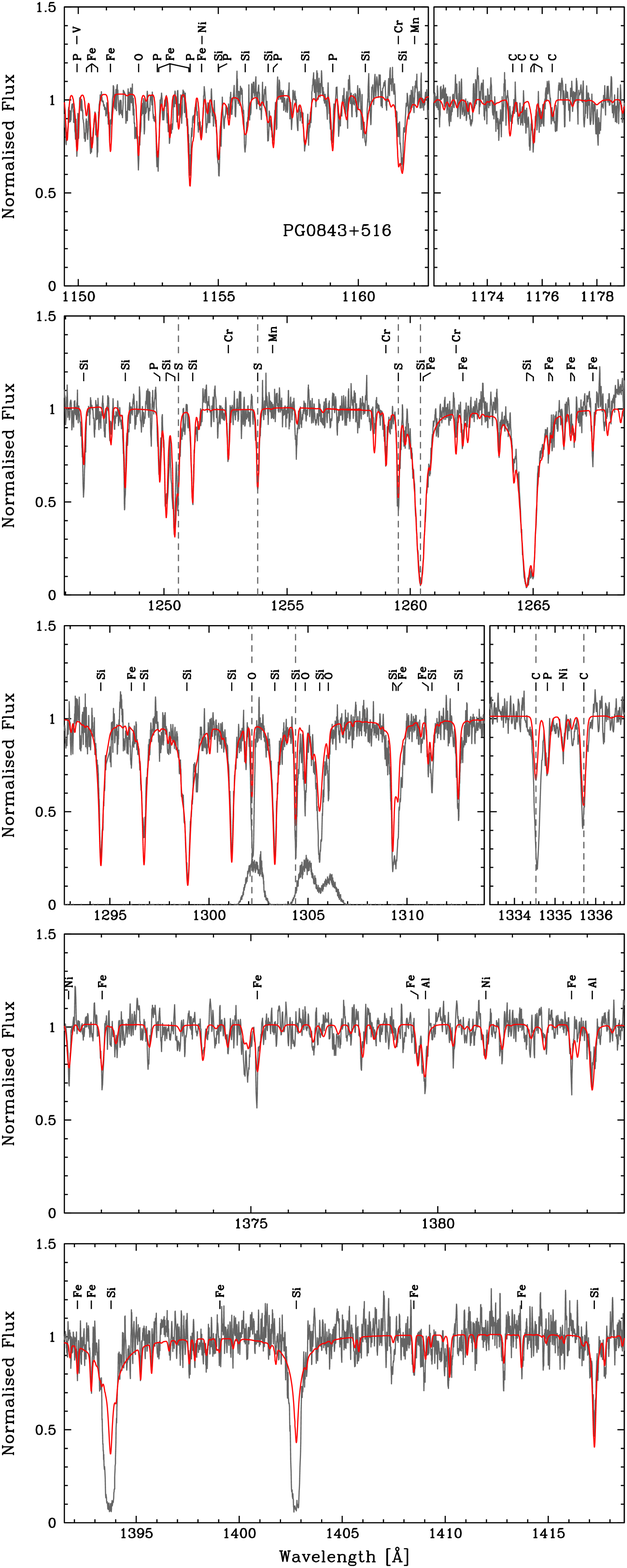}
\includegraphics[width=\columnwidth]{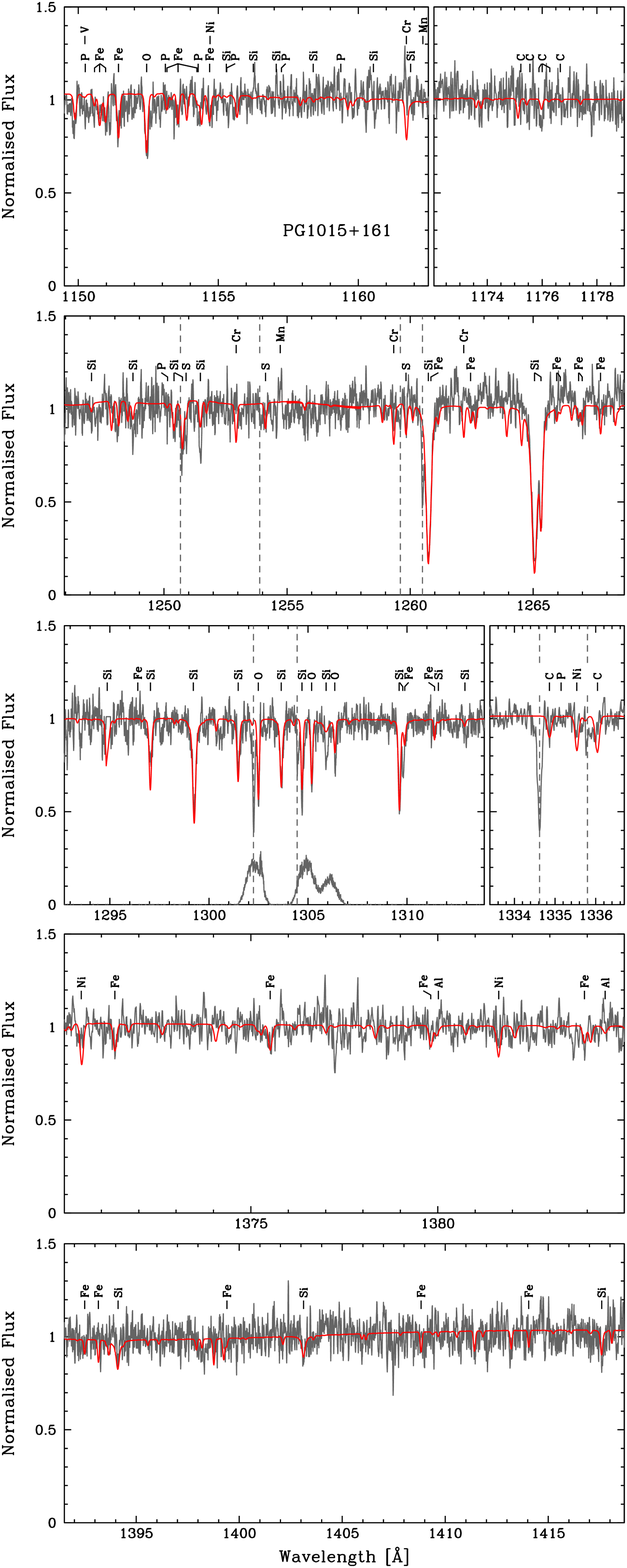}
\caption{\label{f-fit1} The normalised COS spectra (black) of
  PG\,0843+516 (left) and PG\,1015+161 (right), along with our
  best-fit models (red). Interstellar absorption features are
  indicated by vertical gray dashed lines. The interstellar lines in
  PG\,1015+161 are blue-shifted with respect to the photospheric
  features by 57\,\kms, in PG\,0843+516 this shift is
  $\le7$\,\kms. Airglow of \Ion{O}{i} can cause some contamination of
  the $1302-1306$\,\AA\ region. An illustrative airglow emission
  spectrum (arbitrarily scaled in flux) is shown. The strong
  \Lines{Si}{iv}{1394,1403} doublet seen in the COS spectrum of
  PG\,0843+516 is not of photospheric, but circumstellar origin
  (Sect.\,~\ref{s-hotgas}).}
\end{figure*}
\begin{figure*}
\includegraphics[width=\columnwidth]{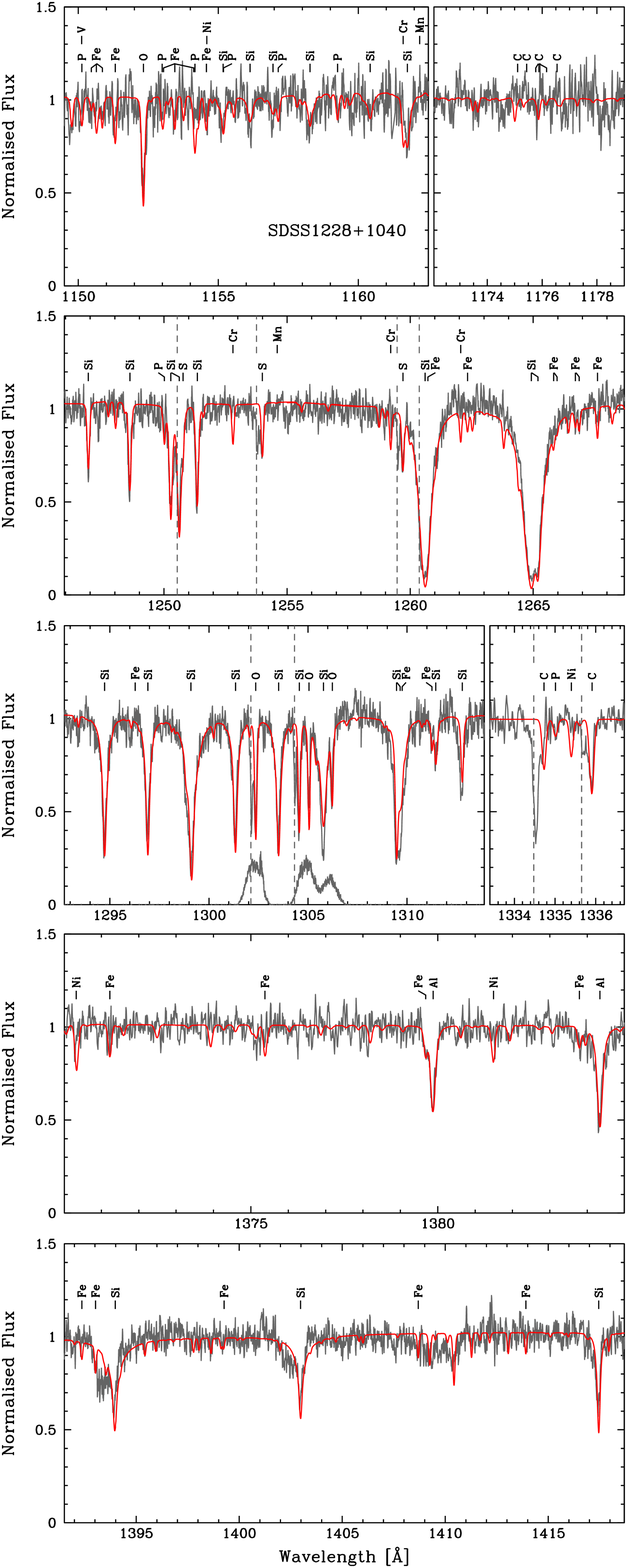}
\includegraphics[width=\columnwidth]{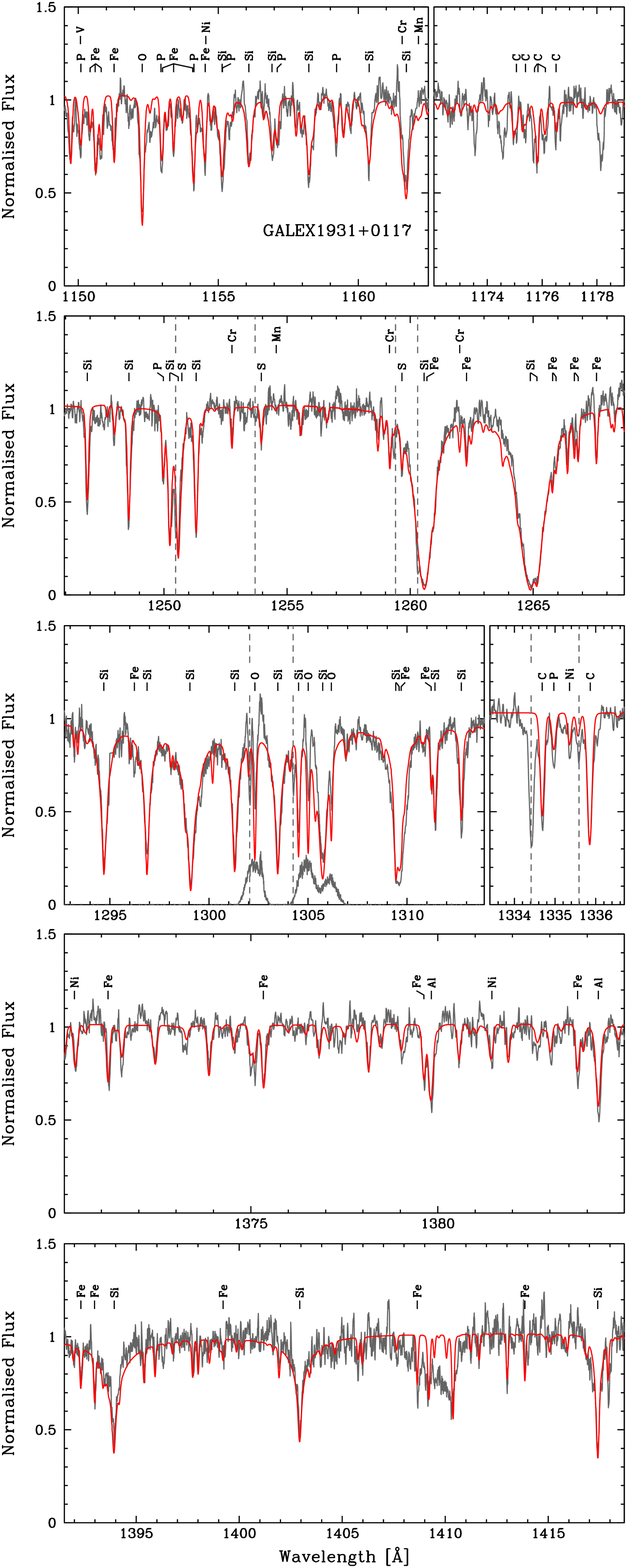}
\caption{\label{f-fit2} Same as Fig.\,\ref{f-fit1}, but for
  SDSS\,1228+1040 (left) and GALEX\,1931+0117 (right).  The
  interstellar lines in their spectra are are blue-shifted with
  respect to the photospheric features by 36\,\kms and 61\,\kms,
  respectively. The strong absorption band seen near 1410\,\AA\ in the
  COS spectrum of GALEX\,1931+0117 is thought to be related to an
  autoionisation line of \Ion{Si}{ii} or to a resonance feature in the
  photoionisation cross section (Sect.\,\ref{s-silicon}). The same
  features is seen, though much weaker, in SDSS\,1228+1040. The
  \Lines{Si}{iv}{1394,1403} doublet in SDSS\,1228+1040 shows
  additional absorption, blue-shifted with respect to the photospheric
  features, which is of circumstellar origin (Sect.\,~\ref{s-hotgas}).}
\end{figure*}

\section{Atmosphere models}

\subsection{\label{s-teff_logg}Effective temperature and surface gravity}

All observed \textit{HST}/COS and optical spectra were analysed with
theoretical model atmospheres using input physics as described in
\citet{koester10-1}, and including the Lyman and Balmer line profiles
of \citet{tremblay+bergeron09-1}. We used a fine grid of models
spanning the range of temperatures and surface gravities found for the
four targets by previous studies (Table~\ref{t-parameters}) and
determined the best-fit parameter by minimising $\chi^2$, using the
very good relative flux calibration as an additional constraint. The
errors reported in Sect.\,\ref{s-notes} are statistical only and do
not include systematic effects of observation, reduction, or
models. More realistic errors can be estimated from a comparison with
the other measurements in the literature, which used similar models,
but optical spectra. Table~\ref{t-parameters} suggests a systematic
trend for somewhat lower temperatures derived from the ultraviolet
data when compared to the values based on optical spectroscopy. A
similar trend is seen for DA white dwarfs with $\Teff\sim20\,000$\,K
in \citet{lajoie+bergeron07-1}, who compared the effective
temperatures derived from optical and (\textit{International
  Ultraviolet Explorer\ }) ultraviolet spectroscopy. We carried out a
range of test calculations to explore the effect of these systematic
uncertainties in $\Teff$ and $\log g$ on the derived metal abundances
(Sect.~\ref{s-abundances}). The abundances and mass fluxes do not
change by more than $\simeq0.1$\,dex, which is less than the typical
uncertainty of our fits, and the abundance ratios vary by much
less. Hence, the discussion in Sect.~\ref{s-debrisnature} and
\ref{s-mdot} is not affected by the systematic uncertainties in
$\Teff$ and $\log g$.

Finally, to assess the possible effect that the presence of
metals has on the effective temperature and surface gravity, we
computed a small grid of models for the two most metal-polluted stars
(PG\,0843+516, GALEX\,1931+0117), including metals at the abundances
determined in Sect.\,\ref{s-abundances}, and re-fitted the
\textit{HST}/COS spectra. For both stars, the best-fit $\Teff$ and
$\log g$ did not change significantly, and we therefore adopted the
atmospheric parameters from the pure-hydrogen fits for all four
targets.

\begin{table}
\caption{\label{t-idlines} List of major line features used for the
  abundance determinations and upper limits. Because of the different
  wavelength ranges of the available spectra not all lines could be
  used for all four stars.}
\begin{tabular}{ll}
\hline
Ion     &  Vacuum wavelengths [\AA]                              \\
 \hline
\Ion{C}{ii}  & 1334.530,1335.660,1335.708 \\
\Ion{C}{iii} & 1174.930,1175.260,1175.590,1175.710,1175.987,1176.370 \\
\Ion{N}{i}   & 1199.550,1200.220,1200.710 \\
\Ion{O}{i}   & 1152.150,1302.170,1304.860,1306.030 \\
\Ion{Mg}{ii} & 1239.925,1240.395,1367.257,1367.708,1369.423, \\
             & 4482.383,4482.407,4482.583 \\
\Ion{Al}{ii} & 1670.787,1719.442,1724.922,1724.982,1760.106,1761.977, \\
             & 1763.869,1763.952,1765.816 \\
\Ion{Al}{iii}& 1379.670,1384.132,1605.766,1611.873 \\
\Ion{Si}{ii} & 1190.416,1193.292,1194.500,1197.394,1246.740,1248.426, \\
             & 1250.091,1250.436,1251.164,1260.422,1264.738,1265.002, \\
             & 1304.370,1305.592,1309.276,1309.453,1311.256,1346.884, \\
             & 1348.543,1350.072,1350.516,1350.656,1352.635,1353.721, \\
             & 1526.707,1533.431,3854.758,3857.112,3863.690,4129.219, \\
             & 4132.059,5042.430,5057.394,6348.864,6373.132 \\
\Ion{Si}{iii}  & 1140.546,1141.579,1142.285,1144.309,1144.959,1154.998, \\
               & 1155.959,1156.782,1158.101,1160.252,1161.579,1206.500, \\
               & 1206.555,1294.545,1296.726,1298.892,1301.149,1303.323, \\
               & 1312.591,1341.458,1342.389,1365.253,1417.237 \\
\Ion{Si}{iv}   & 1393.775,1402.770 \\
\Ion{P}{ii}    & 1149.958,1152.818,1153.995,1155.014,1156.970,1159.086, \\
               & 1249.830,1452.900,1532.533,1535.923,1536.416,1542.304, \\
               & 1543.133,1543.631 \\
\Ion{P}{iii}   & 1334.813,1344.326 \\
\Ion{S}{ii}    & 1250.584,1253.811,1259.519 \\
\Ion{S}{iii}   & 1194.041,1194.433 \\
\Ion{Ca}{ii}   & 1169.029, 1169.198,1341.890,3737.965,3934.777 \\
\Ion{Sc}{ii}   & 1418.773,1418.793 \\
\Ion{Ti}{iii}  & 1298.633,1298.697,1298.996,1327.603 \\
\Ion{V}{iii}   & 1148.465,1149.945,1149.945 \\
\Ion{Cr}{iii}  & 1136.669,1146.342,1247.846,1252.616,1259.018,1261.865, \\
               & 1263.611 \\
\Ion{Mn}{ii}   & 1162.015,1188.505,1192.316,1192.330,1197.184,1199.391, \\
               & 1201.118,1233.956,1254.410 \\
\Ion{Mn}{iii}  & 1174.809,1177.478,1179.851,1183.308,1183.863,1183.880 \\
\Ion{Fe}{ii/iii} &  many weak lines, individually recognisable 1140-1152\\
\Ion{Ni}{ii}    & 1317.217,1335.201,1370.123,1381.286,1393.324,1411.065 \\
\hline
\end{tabular}
\medskip
\end{table}

\subsection{\label{s-abundances} Metal abundances}

The COS spectra of the four white dwarfs contain a multitude of
absorption lines from a range of elements. GALEX\,1931+0117 has the
richest absorption spectrum, in which we securely identified
transitions of nine elements (C, O, Al, Si, P, S, Cr, Fe, Ni), and we
included those metals in the abundance analysis of all four
targets. We also include in the analysis N, Na, Ti, V, Mn, which have
moderately strong transitions in the wavelength range covered by the
COS observations, but that were not detected. All metals were fully
included in the calculation of the equation of state.

Synthetic spectra were calculated adopting the atmospheric parameters
determined in Sect.\,\ref{s-teff_logg}, and including approximately
2500 metal lines. The basic source of atomic line data (wavelengths,
excitation energies, transition probabilities $\log$\,gf, Stark
broadening constant $\Gamma_4$) was VALD (Vienna Atomic Line
Database), which is described in \cite{piskunovetal95-1},
\cite{ryabchikovaetal97-1}, and \cite{kupkaetal99-1,
  kupkaetal00-1}. The ion \Ion{Si}{ii} has a large number of lines in
the ultraviolet, and we noted a significant scatter in the abundances
derived from different lines. Replacing the $\log$~gf values from VALD
values with those from the NIST (National Institute of Standards)
database, which differ for some lines by up to 0.3\,dex, leads to more
consistent results. Nevertheless, the situation for this ion is not
satisfactory (Sect.\,\ref{s-silicon}), and we have consulted a number of original
sources in the literature \citep{lanz+artru85-1, nahar98-1,
  bautistaetal09-1} during the compilation of the most reliable atomic
data.

The abundances were varied until a satisfactory fit, as judged by
visual inspection, was achieved for each element. We then changed the
abundances in several steps of 0.1~-~0.2 dex, until the fit was
clearly worse. The resulting difference was used as a conservative
estimate for the abundance error, or for an upper limit if no line was
identified. Table~\ref{t-idlines} lists the lines used in this
procedure, although not all lines could be used for all four stars.
The best-fit models to the COS observations are illustrated in
Figs.\,\ref{f-fit1}~and~\ref{f-fit2}, and the metal abundances of the
four white dwarfs are given in Table\,\ref{t-abundances} (along with
the previous abundance studies were carried out for GALEX\,1931+0117,
\citealt{vennesetal11-1, melisetal11-1}). Notably, upper limits for N
were always larger than solar relative to C.  For Na, Ti, V, Mn (and
additionally Ca in PG\,0843+516 as well as Ca, Al, P, S, Ni in
PG\,1015+161) the upper limits were larger than solar relative to
Si. We have used these (solar) values in the models, but it did not
change the atmosphere structure and the results for the detected
elements.

\subsubsection{Interstellar line absorption and airglow}
In all objects interstellar absorption is visible in the resonance
lines of \Ion{C}{ii}, \Ion{N}{i}, \Ion{O}{i}, \Ion{Si}{ii}, and
\Ion{S}{ii}. In SDSS\,1228+1040, PG\,1015+161, and GALEX\,1931+0117
the interstellar absorption lines are shifted blue-wards with respect
to the photospheric lines by velocities of $v = 57$, $36$, and
$61$~\kms, respectively. In PG\,0843+516, $|v| < 7$ \kms, and the
interstellar lines are not fully separated from the photospheric
features. However, the presence of some interstellar absorption is
obvious from the line ratio of
\Line{C}{ii}{1334.5}/\Line{C}{ii}{1335.7} (Fig.\,~\ref{f-fit1} \&
\ref{f-fit2}). Because the latter line originates from a level only
0.008~eV above the real ground state, it is equally populated in a
stellar photosphere, but not in the interstellar medium, where the
blue component is much stronger in spite of a lower transition
probability. Nevertheless, the abundances of C, O, Si, and S are
robust, as a sufficient number of excited transitions are present in
the photospheric spectrum (Table\,\ref{t-idlines}).

The COS pipeline does not correct for airglow emission. Therefore, the
reduced COS spectra can contain geocoronal lines of \Lines{O}{i}{1302,
  1305, 1306} whose intensity, and, to a lesser extent, profile shape,
vary as a function of \textit{HST}'s orbital day/night, and weakly with
the Earth-limb angle. Airglow is clearly seen in the spectrum of
GALEX\,1931+0117 (Fig.\,\ref{f-fit2}, right panel), which affects the
fit to the photospheric \Ion{O}{i} and \Ion{Si}{ii} lines in this
region. For Si, this is a minor problem as there are many additional
lines of \Ion{Si}{ii-iv}. For O, another strong line in the COS
spectra is \Line{O}{i}{1152}.

\subsubsection{Silicon}
\label{s-silicon}
We notice relatively large differences of the silicon abundance
determined from optical versus ultraviolet spectra in SDSS\,1228+1040\ and
GALEX\,1931+0117, for the latter also the oxygen abundances show this
difference. There are at least three possible explanations:

\textit{Uncertain atomic data.} This is a perennial problem, as there
are many, and large differences in various compilations of atomic
data. The \Ion{O}{i} resonance lines in GALEX\,1931+0117 are perturbed
by airglow, interstellar absorption and overlapping \Ion{Si}{ii} lines
(see above), and the ultraviolet abundance determination rests largely
on one excited line at 1152.1~\AA. Similarly, the optical O abundance
is measured only from the \Lines{O}{i}{7777} triplet
\citep{vennesetal10-1, melisetal11-1}. However, our abundance
measurements for Si use many lines in the ultraviolet. In the recent
compilation by \cite{bautistaetal09-1} the authors combined several
different computational methods, previous theoretical calculations by
other authors, and experimental data into a ``recommended'' value for
$\log$~gf. These values agree fairly well with the ultraviolet data
from NIST that we have used. However, for the five optical lines they
consider, the values are $0.25-0.30$\,dex smaller, though with errors
as large as 0.3\,dex. Using these values would {\em increase} the
abundance determined from optical spectra, contrary to what would be
needed for a more consistent solution. In addition, in a recent
analysis of ultraviolet spectra for the DBZ star GD\,40,
\citet{juraetal12-1} find a discrepancy between optical and
ultraviolet abundances for Si of the same size, but in opposite
direction - the abundances are smaller for the optical
determinations. Since that study used the same models and atomic data
as the one presented here, there is no indication that the atomic data
are behind this discrepancy.

\begin{table*}
\caption{\label{t-abundances} Element number abundances
  $\log[\mathrm{Z/H}]$ and limits determined from the analysis of our
  \textit{HST}/COS and optical spectra. For Si, we separately report
  the abundances determined from the optical data alone (``opt'', see
  Sect.\,\ref{s-silicon}), and for Ca and Mg, we report both the
  abundances from a homogenous and a stratified atmosphere (``strat'',
  see Sect.\,\ref{s-diffusion}), the latter ones are preferred as the
  strongest lines for both elements are observed are in the
  optical. For GALEX\,1931+0117, we also list the results of
  \citet{vennesetal11-1} and \citet{melisetal11-1}.}
\begin{center}
 \begin{tabular}{lrrr@{\hspace*{10ex}}rrr}
 \hline
 Element     & PG\,0843+516   &  PG\,1015+161  &  SDSS\,1228+1040 &
 \multicolumn{3}{c}{~\dotfill~~~~GALEX\,1931+0117~~~~\dotfill~}   \\
             & \multicolumn{4}{c}{~\dotfill~~~~this
   paper~~~~\dotfill~} & Vennes et al. & Melis et al.\\
 \hline
 C           & $-7.30\pm0.30$ & $<-8.00$       & $-7.50\pm0.20$ & $-6.80\pm0.30$ & $<-4.15$        & $<-4.85$       \\ 
 O           & $-5.00\pm0.30$ & $-5.50\pm0.20$ & $-4.55\pm0.20$ & $-4.10\pm0.30$ & $-3.62\pm0.05$  & $-3.68\pm0.10$ \\
 Mg          & $-4.90\pm0.20$ & $-5.30\pm0.20$ & $-5.10\pm0.20$ &                & $-4.42\pm0.06$  & $-4.10\pm0.10$ \\
 Mg~(strat)  & $-5.00\pm0.20$ & $-5.30\pm0.20$ & $-5.20\pm0.20$ &                &                 &                \\
 Al          & $-6.50\pm0.20$ &                & $-5.75\pm0.20$ & $-6.20\pm0.20$ &                 &                \\
 Si          & $-5.20\pm0.20$ & $-6.40\pm0.20$ & $-5.20\pm0.20$ & $-4.75\pm0.20$ &                 &                \\
 Si~(opt)    &                &                & $-4.70\pm0.20$ &                & $-4.24\pm0.07$  & $-4.35\pm0.11$ \\ 
 P           & $-6.60\pm0.20$ &                & $<-7.30$       & $-7.00\pm0.30$ &                 &                \\
 S           & $-5.50\pm0.30$ &                & $<-6.20$       & $-6.60\pm0.20$ &                 &                \\
 Ca          &                & $-6.30\pm0.20$ & $-5.70\pm0.20$ &                & $-6.11\pm0.04$  & $-5.83\pm0.10$ \\ 
 Ca~(strat)  &                & $-6.45\pm0.20$ & $-5.94\pm0.20$ &                &                 &                \\
 Cr          & $-5.80\pm0.30$ & $<-5.80$       &     $<-6.00$   & $-6.10\pm0.30$ &                 & $-5.92\pm0.14$ \\
 Mn          &                &                &                &                &                 & $-6.26\pm0.15$ \\ 
 Fe          & $-4.60\pm0.20$ & $-5.50\pm0.30$ & $-5.20\pm0.30$ & $-4.50\pm0.30$ & $ -4.43\pm0.09$ & $-4.10\pm0.10$ \\
 Ni          & $-6.30\pm0.30$ &                & $<-6.50$       & $-6.70\pm0.30$ &                 & $<-5.60$       \\
\hline
\end{tabular}
\end{center}
\end{table*}

\textit{Abundance stratification.}  Contrary to DB stars like GD\,40
at similar temperatures, there are no convection zones in the
atmospheres and envelopes of our four objects, which would act as a
homogeneously mixed reservoir in the accretion/diffusion
scenario. Assuming a steady state between the two processes, we thus
expect a stratified abundance configuration. Whether this can explain
the observations will be studied in Sect.\,\ref{s-diffusion}.

\textit{Genuine variation of the accretion rates.} As will also be
discussed in the next section, the time scales for diffusion in these
atmospheres are of the order of days. If the accretion rate is not
constant the observed abundances may change on the same short time
scales. Given that the COS and ground-based observations that we
analysed were taken months to years apart, such variations can not be
excluded. Noticeable variations of the \Ion{Ca}{ii} equivalent widths
in the debris disc white dwarf G29-38 were reported by
\citet{vonhippeletal07-2}. However, a similar study on the same star
by \citet{debes+lopez-morales08-1} did not find any variations in the
line strengths. Thus, the current evidence for accretion rate
variations on time scales of months to years is ambiguous, and a
second-epoch COS observations of the stars studied here would be
desirable.

We also noticed an unidentified absorption feature between 1400 and
1410~\AA, with a strength roughly correlated with the Si abundances. Such
a feature has been discussed in the literature and related to an
autoionisation line of \Ion{Si}{ii} or to a resonance feature in the
photoionisation cross section \citep{artru+lanz87-1, lanzetal96-1}. We
have tested such a hypothetical line with their data for the
oscillator strength and line width data. However, the width ($\approx
80$\,\AA) is much too broad to lead to visible features in the
spectrum. We have also included the \Ion{Si}{ii} photoionisation cross
sections from the Opacity Project \citep{seatonetal94-1}, which indeed
show a resonance maximum in this spectral region. But again, the Si
abundance is too small to let this feature show up in the spectrum.

Our model uses the six \Ion{Si}{ii} lines at 1403.8, 1404.2, 1404.5,
1409.1, 1409.9, and 1410.2~\AA\ in this range
(Table\,\ref{t-idlines}). The first two have the source ``guess'' in
VALD, the first three have no entry in NIST, and the $\log$~gf values
of the strongest line (1410.2\,\AA) differ by $\approx 0.8$\,dex
between the two databases. The upper levels of the transitions have a
parent configuration belonging to the second ionisation limit of
\Ion{Si}{ii}. They are still $\approx0.7$\,eV below the first
ionisation limit and thus not strictly auto-ionising. However, the
broadening may well be underestimated by our simple approximation
formulae. In summary, the atomic data of the lines in the region are
very uncertain and may be the explanation for the broad
feature. However, with the present data we cannot prove that
hypothesis.

Finally, we note that the \Lines{Si}{iv}{1394,1403} doublet in
PG\,0843+516 is very poorly fit by our atmosphere model
(Fig.\,\ref{f-fit1}). A weaker additional \Lines{Si}{iv}{1394,1403}
absorption is also seen in the spectrum of SDSS\,1228+1040
(Fig.\,\ref{f-fit2}). We interpret this as evidence for absorption by
hot gas close to the white dwarf, see the discussion in
Sect.\,\ref{s-hotgas}.

\begin{table*}
 \centering
 \caption{Diffusion fluxes $\rho X v\,[\mathrm{g\,s^{-1}}]$ within the
   white dwarf atmospheres, which are equal to the rates at which
   planetary debris material is accreted. $\Sigma$ gives the sum of
   the accretions rates of all detected elements (i.e. not
     including those with upper limits). For GALEX\,1931+0117, we
     adopt in our calculation the Mg and Ca abundances of
     \citet{vennesetal11-1} and the Mn abundance of
     \citet{melisetal11-1} \label{t-fluxes}.}
 \begin{tabular}{lrrrr}
 \hline
 Element    & \multicolumn{1}{c}{PG\,0843+516}    &  \multicolumn{1}{c}{PG\,1015+161}
            & \multicolumn{1}{c}{SDSS\,1228+1040} &  \multicolumn{1}{c}{GALEX\,1931+0117}    \\
 \hline
 C          &  $1.66\times10^{5}$ &  $4.65\times10^{4}$  & $1.25\times10^{5}$  & $4.57\times10^{5}$  \\
 O          &  $9.27\times10^{7}$ &  $3.78\times10^{7}$  & $2.70\times10^{8}$  & $5.61\times10^{8}$  \\
 Mg         &  $4.47\times10^{7}$ &  $2.66\times10^{7}$  & $3.21\times10^{7}$  & $1.47\times10^{8}$  \\
 Al         &  $2.09\times10^{6}$ &                      & $1.18\times10^{7}$  & $3.08\times10^{6}$  \\
 Si         &  $4.77\times10^{7}$ &  $3.64\times10^{6}$  & $4.80\times10^{7}$  & $9.93\times10^{7}$  \\
 P          &  $2.44\times10^{6}$ &                      & $<5.24\times10^{5}$ & $7.57\times10^{5}$  \\ 
 S          &  $3.92\times10^{7}$ &                      & $<9.46\times10^{6}$ & $2.64\times10^{6}$  \\
 Ca         &                     &  $4.84\times10^{6}$  & $1.57\times10^{7}$  & $8.10\times10^{6}$  \\
 Cr         &  $3.81\times10^{7}$ &  $<3.85\times10^{7}$ & $<2.29\times10^{7}$ & $1.37\times10^{7}$  \\
 Mn         &                     &                      &                     & $1.06\times10^{7}$  \\
 Fe         &  $7.11\times10^{8}$ &  $9.50\times10^{7}$  & $1.72\times10^{8}$  & $6.45\times10^{8}$  \\
 Ni         &  $1.66\times10^{7}$ &                      & $<9.98\times10^{6}$ & $4.71\times10^{6}$  \\
\noalign{\smallskip}
$\Sigma$    &  $1.02\times10^{9}$ &  $1.68\times10^{8}$  & $5.61\times10^{8}$  & $1.50\times10^{9}$  \\
\hline
\end{tabular}

\end{table*}

\subsubsection{Diffusion and stratified atmosphere models}
\label{s-diffusion}
In the absence of a convection zone there is no deep homogenous
reservoir in our DAZ sample, and therefore there is no straightforward
definition of diffusion time scales. Adopting the usual definition,
i.e. dividing the mass of some element above a layer in the envelope
or atmosphere of the star by the diffusion flux, results in diffusion
time scales that strongly depend on the chosen
layer. \cite{koester+wilken06-1} and \cite{koester09-1} defined the
Rosseland optical depth $\tau=5$ as the ``standard'' layer, assuming
that no trace of any heavy element below this would be seen in a
spectrum.

However, a more consistent way to determine the abundances in the
accreted material, which is the quantity ultimately desired, is the
assumption of a steady state between accretion and diffusion
throughout the whole atmosphere. At Rosseland optical depth $\tau =
2/3$, and typical conditions for the observed ultraviolet spectra, the
diffusion times in the four white dwarfs analysed here are $\simeq0.4$
to four days. Assuming that the accretion rate does not vary over such
time scales, we can use the condition of constant flow of an element
with mass fraction $X(\tau)$
\begin{equation}
\rho X v = \mbox{const}
\end{equation}
with $\rho$ and $v$ the mass density and the diffusion velocity of
this element. $\rho$ and $v$ are known from the atmosphere model and
diffusion calculations, and $X(\tau=2/3)$ is derived from the spectral
analysis. This determines the diffusion flux at $\tau = 2/3$. In
steady state, as it is the case for the DAZ analysed here, the
diffusion flux is constant throughout the atmosphere, and is equal to
the accretion rate polluting the atmosphere. The constant diffusion
flux then in turn allows the determination of the abundance
stratification $X(\tau)$ \citep[see also][for a thorough
  discussion]{vennesetal11-1}.

We calculated new stratified models and synthetic spectra for all
objects, using the steady state condition and the abundances (at $\tau
= 2/3$) from Table~\ref{t-abundances}. The resulting spectra are
almost indistinguishable from those of the homogeneous atmospheres; the
only exception are small increases of the optical \Ion{Mg}{ii} and
\Ion{Ca}{ii} line strengths.  The small change can easily be explained
by the structure of the stratified atmosphere. In these models $\rho
\,v$ increases with depth, and consequently the abundance
decreases. On the other hand a monochromatic optical depth of $\approx
2/3$ is reached in the ultraviolet near Rosseland optical depth of
$\tau_\mathrm{Ross}\simeq2/3$, while it is reached at
$\tau_\mathrm{Ross} \approx 0.15$ for $\lambda = 4480$~\AA, i.e.
higher in the atmosphere, where the abundance is correspondingly higher.

For PG\,0843+516, PG\,1015+161, and SDSS\,1228+1040, the Ca and Mg
abundances were obtained from the optical data (Sect.\,\ref{s-optobs})
and our models. We have iterated them by fitting to stratified models
(denoted with ``strat'' in Table~\ref{t-abundances}). For
  GALEX\,1931+0117, we adopted the photospheric Mg an Ca abundances of
  \citet{vennesetal11-1} and the Mn abundance of \citet{melisetal11-1}
  to calculate the corresponding diffusion fluxes.

As a result we have to conclude that diffusion and a stratified
abundance structure lead only to minor adjustments of the abundances
that cannot explain the large discrepancy between optical and
ultraviolet determinations for silicon. There is, however, an
important caveat to this conclusion. Our diffusion calculations use
only the surface gravity (and as a minor effect the temperature
gradient for thermal diffusion) as driving
force. \cite{chayer+dupuis10-1} have recently demonstrated that for
silicon, radiative levitation can lead to a negative effective gravity
and support the atoms in the outer layers of the atmosphere against
diffusion. They only published detailed data for a DAZ model with
20\,000\,K and $\log g=8.00$, and in their model only abundances
smaller than $\log\mathrm{[Si/H]}=-8.0$ are really supported, because
the lines saturate at higher abundances, effectively reducing the
radiative support. However, it is quite feasible that even if the
atoms are not totally supported, the diffusion velocity would be
smaller, changing the abundance gradient. The answer to this puzzle
will have to await similar, detailed models for a variety of stellar
parameters and heavy elements that can be tested against the large
range of Si abundances found in our snapshot survey (G\"ansicke et
al. in prep).

Other points worth mentioning are that the determination of an
effective ion charge with the simple pressure ionisation description
of \citet{paquetteetal86-1} is not appropriate in the absence of deep
convection. We have used the usual Saha equation (with a small
lowering of the ionisation potential from non-ideal interactions) to
determine the abundances of different ions from an element. The
diffusion velocity is then calculated as a weighted average of the
ionisation stages. This procedure was already used in
\citet{koester+wilken06-1} and \citet{koester09-1} for the models
without or with only a shallow convection zone, although not
explicitly stated in those papers.
New in our present calculation is the consideration of neutral
particles, following the discussion and methods outlined in
\citet{vennesetal11-2}. 

\begin{figure}
\centerline{\includegraphics[angle=270,width=\columnwidth]{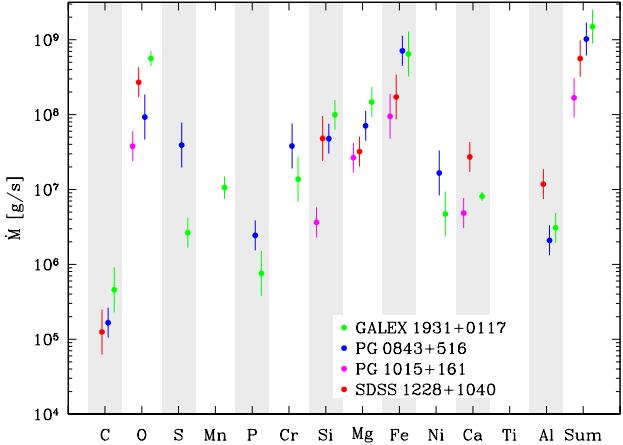}}
\caption{\label{f-mdot} Accretion rates of the elements detected in
  our four targets. Their  sum is given in the right-most column.}
\end{figure}

The main results of our calculations are the diffusion fluxes, $X \rho
v$, for each element, which are assumed (in steady state) to be the
abundances of the accreted matter. These are summarised for the four
objects in Table~\ref{t-fluxes}.   The total
diffusion fluxes (\,=\,accretion rates) are obtained by multiplying
these fluxes with $4 \pi R_\mathrm{wd}^2$, where we used the cooling
tracks of \cite{wood95-1} to obtain the white dwarf radii from $\Teff$
and $\log g$. The mass fluxes (\,=\,accretion rates) of the individual
elements, as well as their sum, are shown in Fig.\,\ref{f-mdot}
and discussed in Sect.~\ref{s-mdot}. The number abundances of
the circumstellar debris are then calculated from the diffusion fluxes
via
\begin{equation}
\mathrm{\frac{N(X)}{N(Si)}=\frac{\dot M(X)}{\dot M(Si)}\frac{A(Si)}{A(X)}}
\end{equation}
where A is the atomic mass.  The implications that these abundances
have on the nature and origin of the circumstellar debris are
discussed in detail in Sect.\,\ref{s-debrisnature}.

\section{Notes on individual white dwarfs}
\label{s-notes}

In the following sections, we give a brief overview of previous work
on the four white dwarfs that we have analysed, as well as a summary
of the key results of our observations.

\subsection{PG\,0843+516}
PG\,0843+516 was identified as a DA white dwarf in the Palomar-Green
Survey \citep{greenetal86-1}, and \citet{liebertetal05-1} obtained
$\Teff=23\,870\pm392$\,K, $\log g=7.90\pm0.05$ from the analysis of a
high-quality optical spectrum. The best fit to our \textit{HST} data
was $\Teff=23\,095 \pm 230$\,K, $\log g=8.17\pm0.06$.  Our COS spectrum
reveals PG\,0843+516 to be an extremely polluted DAZ white dwarf
(Fig.\,\ref{f-cos} \& \ref{f-fit1}), with an accretion rate of
$\simeq10^9\,\mathrm{g\,s^{-1}}$, placing it head-to-head with GALEX\,1931+0117
(Sect.\,\ref{s-mdot}). We identified in the COS spectrum photospheric
absorption lines of C, O, Al, Si, P, S, Fe, Cr, and Ni, plus Mg in the
optical WHT spectrum, the second largest set of elements detected in a
DAZ white dwarf.
The fact that the metal pollution of PG\,0843+516 went unnoticed in
the published high-quality intermediate resolution spectroscopy
underlines the strength of our ultraviolet survey for young and
relatively warm white dwarfs accreting planetary debris. We note that
\citet{xu+jura12-1} recently detected infrared flux excess at
PG\,0843+516 in an analysis of archival \textit{Spitzer} data, making
this the second white dwarf (after G29-38,
\citealt{zuckerman+becklin87-1, koesteretal97-1}) where circumstellar
dust was found without prior knowledge of photospheric metal
pollution.

\subsection{PG\,1015+161}
\label{s-pg1015}
PG\,1015+161 is another DA white dwarf discovered in the Palomar-Green
Survey \citep{greenetal86-1}. \citet{liebertetal05-1} determined
$\Teff=19\,540\pm305$\,K, $\log g = 8.04\pm0.05$ from optical
spectroscopy. High-resolution spectroscopy of PG\,1015+161 was
obtained as part of the SPY project \citep{napiwotzkietal01-1}, from
which \citet{koesteretal09-2} measured $\Teff=19\,948\pm33$\,K and $\log
g=7.925\pm0.006$. Our fit to the HST spectrum gives in $\Teff=19\,200 \pm
180$\,K, $\log g = 8.22 \pm 0.06$. \citet{koesteretal05-2} detected of
a photospheric \Ion{Ca}{ii}\,K absorption line in the SPY data, with a number
abundance $\log\mathrm{[Ca/H]}=-6.3$, which triggered follow-up
observations with \textit{Spitzer} that revealed the presence of
circumstellar dust \citep{juraetal07-1}.
The COS spectrum contains absorption lines of O, Si, and Fe. In
addition to \Ion{Ca}{ii}\,K, we detected \Line{Mg}{ii}{4482} in the SPY
spectrum. PG\,1015+161 has the lowest accretion rate among the four
stars discussed in this paper.

\subsection{SDSS\,1228+1040}
\citet{eisensteinetal06-1} identified this DA white dwarf in Data
Release~4 of the Sloan Digital Sky Survey, and found
$\Teff=22\,125\pm136$\,K, $\log g=8.22\pm0.02$ from a fit to the SDSS
spectrum. \citet{gaensickeetal06-3} discovered double-peaked emission
lines of \Lines{Ca}{ii}{8498,8542,8662} as well as weak \Ion{Fe}{ii}
emission lines and \Line{Mg}{ii}{4482} absorption, and concluded that
SDSS\,1228+1040 accretes from a volatile-depleted gaseous
circumstellar disc. The \Ion{Ca}{ii} lines form in a region
  extending in radius from a few tenths \Rsun\ to $\simeq1.2$\,\Rsun,
  no emission is detected from closer in to the white dwarf (but see
  Sect.~\ref{s-hotgas}). \textit{Spitzer} observations showed that
  SDSS\,1228+1040 also exhibits an infrared excess
  \citep{brinkworthetal09-1}, and that there is a large radial overlap
  between the gaseous and dusty components of the disc. Yet, the
  strong \Ion{Ca}{ii} emission lines require a gas temperature of
  $T\sim4000-6000$\,K (e.g. \citealt{hartmannetal11-1}), substantially
  exceeding the sublimation temperature of the dust. This implies the
  thermal decoupling of the gas and dust, most likely in the form of a
  complex vertical temperature structure, with hotter, optically thin
  gas on top cooler, probably optically thick dust \citep{kinnear11,
    melisetal10-1}. Irradiation from the white dwarf is sufficient to
explain this temperature inversion \citep{kinnear11, melisetal10-1},
but the origin of the gas found at radii larger than the sublimation
radius is unclear, and may be related to relatively fresh disruption
events \citep{gaensickeetal08-1, melisetal10-1} or the intrinsic
evolution of the debris disc \citep{bochkarev+rafikov01-1,
  metzgeretal12-1}. Among the four white dwarfs studied here,
SDSS\,1228+1040 is the only one that exhibits emission lines from a
gaseous disc.

The COS spectrum of SDSS\,1228+1040 contains absorption lines of C, O,
Al, Si, Cr, and Ni.  SDSS\,1228+1040 was observed outside the snapshot
program described in Sect.\,\ref{s-hstobs}, and our COS spectroscopy
extends up to 1790\,\AA, i.e. 360\,\AA\ further than that obtained for
the other three white dwarfs. This extended wavelength range includes
additional strong lines of \Ion{Si}{ii}, \Ion{Al}{ii}, and
\Ion{Al}{iii}, but no further elements. Our high-quality average UVES
spectrum is used to determine the abundances of Mg and Ca, bringing
the total number of detected elements in SDSS\,1228+1040 to eight.

We fitted the SDSS spectrum, finding $\Teff=22\,410\pm175$\,K, $\log g=
8.12 \pm 0.03$, whereas a fit to the ultraviolet spectrum gives
$\Teff=20\,565 \pm 82$\,K, $\log g=8.19 \pm 0.03$. This discrepancy
underlines that, for high-quality data, the uncertainties are
dominated by systematic rather than statistical errors.  As a
compromise we take the weighted mean of the latter two results with
increased errors, $\Teff=20\,900 \pm 900$\,K, $\log g = 8.15 \pm 0.04$.

\subsection{GALEX\,1931+0117}
As part of a spectroscopic identification program of
ultraviolet-excess objects \citet{vennesetal10-1} recently identified
GALEX\,1931+0117 as a nearby ($\simeq55$\,pc) DAZ white dwarf.
\citet{vennesetal10-1} and \citet{melisetal11-1} analysed optical
spectroscopy, and obtained $\Teff = 20\,890\pm120$\,K, $\log
g=7.90{+0.03\atop-0.06}$ and $\Teff = 23\,470\pm300$\,K, $\log
g=7.99\pm0.05$, respectively. Our best-fit parameters from the
\textit{HST}/COS spectrum are $\Teff=21\,200\pm50$\,K, $\log
g=7.91\pm0.02$, consistent with that of \citet{vennesetal10-1} but
somewhat lower than that of
\citet{melisetal11-1}\footnote{\citet{melisetal11-1} discuss the
  discrepancy between their model and the \textit{GALEX} fluxes. From
  their Table\,1, it appears that they did not correct for the
  non-linearity of the \textit{GALEX} detectors for bright
  targets. The corrected \textit{GALEX} magnitudes given by
  \citet{vennesetal10-1} are in good agreement with our best-fit
  model.}.
The VLT/UVES spectroscopy obtained by \citet{vennesetal10-1,
  vennesetal11-1} revealed strong metal lines of O, Mg, Si, Ca, and
Fe, indicating ongoing accretion. \citet{vennesetal10-1} also showed
that the 2MASS $H$- and $K$-band fluxes exceeded those expected from
the white dwarf, and suggested a close brown dwarf or a dusty debris
disc as origin of the accreting material. \citet{debesetal11-1} ruled
out the presence of a sub-stellar companion based on the infrared
fluxes detected by \textit{WISE}, and argued that the white dwarf
accretes from a dusty disc. This was independently confirmed by
VLT/ISAAC near-IR observations obtained by \citet{melisetal11-1}, who
also measured abundances for Cr and Mn.

\begin{figure*}
\includegraphics[angle=270,width=\columnwidth]{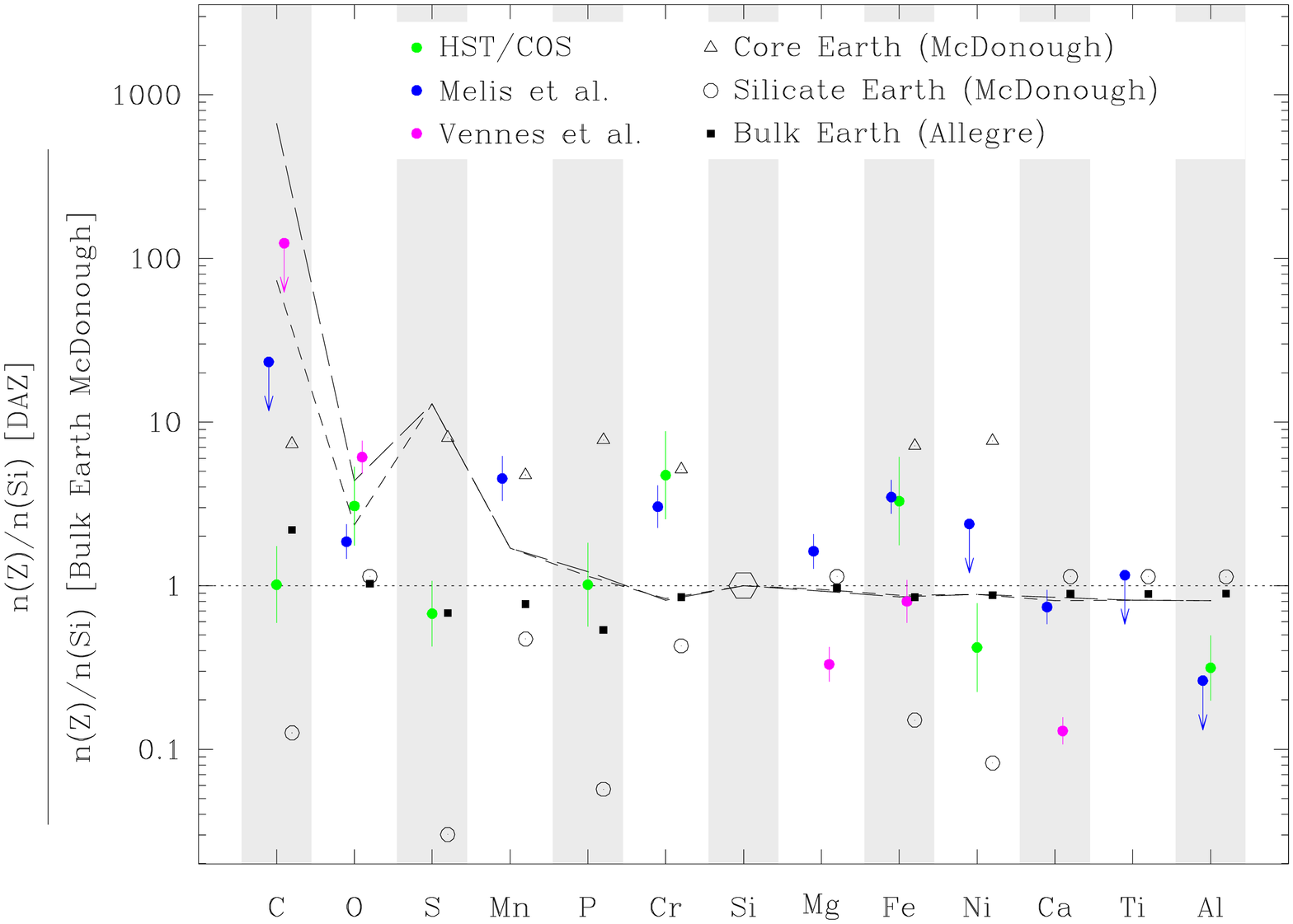}
\includegraphics[angle=270,width=\columnwidth]{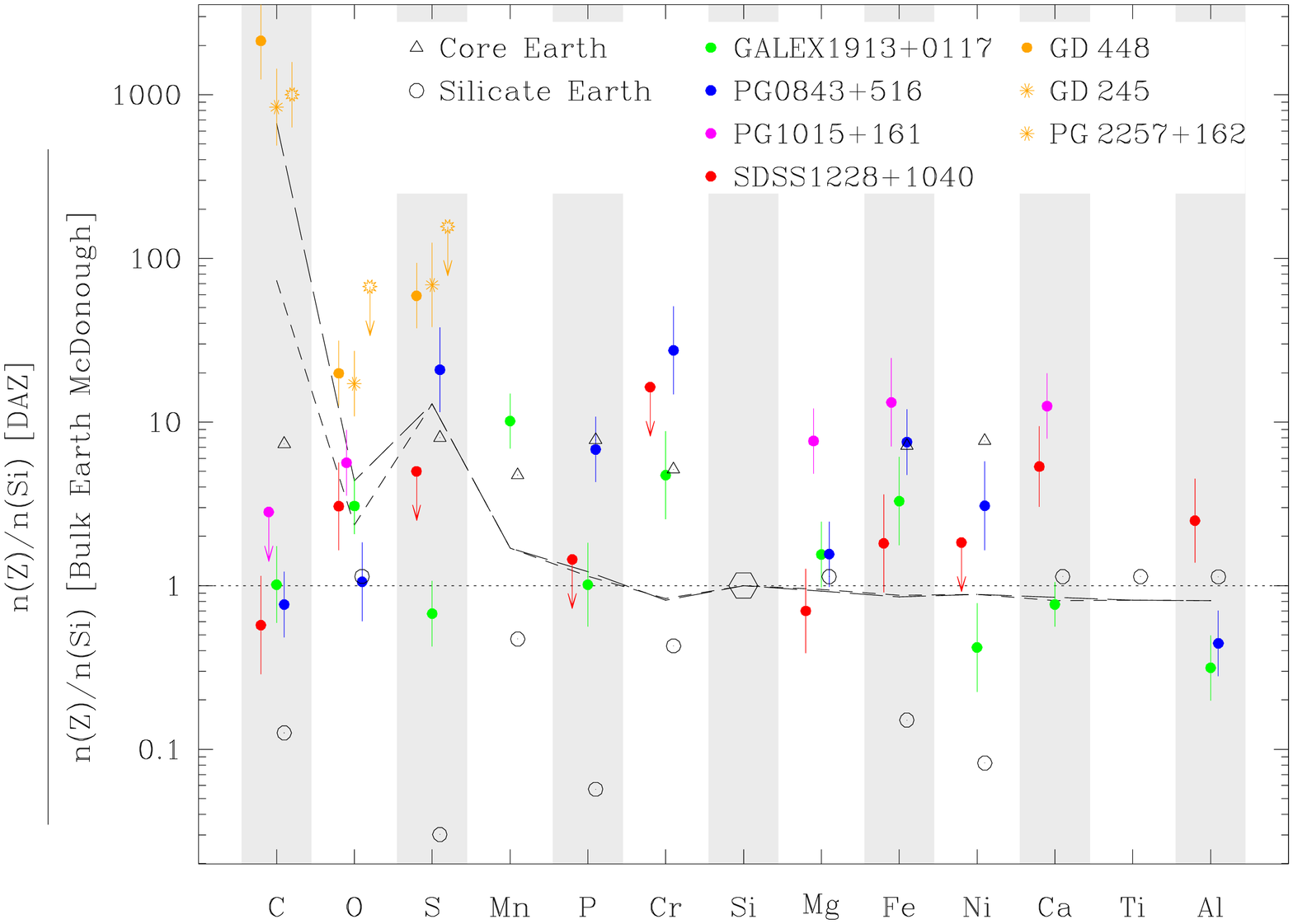}
\caption{\label{f-mm} Heavy element abundances derived for the
  circumstellar debris at the four white dwarf targets
  (see Table\,~\ref{t-fluxes}), relative to Si, and normalised to
  the same ratios of the bulk Earth \citep{mcdonough00-1}. The
  elements are arranged, left to right, in order of increasing
  sublimation temperature \citep{lodders03-1}. \textit{Left panel:}
  abundances for the debris around GALEX\,1931+0117 (green: this
  paper, blue: \citealt{melisetal11-1}, magenta:
  \citealt{vennesetal11-1}). Also shown are the abundance ratios for
  the core Earth (which makes up $\sim1/3$ of the Earth's mass, open
  triangles) and the silicate Earth (i.e crust and mantle, which make
  up $\sim2/3$ of the Earth's mass, open circles). The bulk Earth
  composition of \citet{allegreetal95-1} is shown as solid black
  squares, illustrating the level of uncertainty in the
  (model-dependent) composition of the Earth. The short-dashed line
  shows the abundance ratios of CI chondrites, the long-dashed line
  those corresponding to solar abundances (both from
  \citealt{lodders03-1}). \textit{Right panel:} Metal-to-Si ratios for
  PG\,0843+516 (blue), PG\,1015+161 (magenta), SDSS\,1228+1040 (red),
  and GALEX\,1931+0117 (green). Shown in orange are the abundance
  ratios of three white dwarfs that accrete from the wind of a close
  M-dwarf companion. As expected for the accretion of material with
  near-solar abundances, the volatiles C and S are found to be
  strongly enhanced compared to the four white dwarfs that host debris
  discs of exo-terrestrial material.}
\end{figure*}

Our \textit{HST}/COS spectroscopy provides independent measurements
for O, Si, Cr, and Fe, as well as the first detection of C, Al, P, S,
and Ni, bringing the total number of elements observed in the
photosphere of GALEX\,1931+0117 to 11 (Table\,\ref{t-abundances}). As
discussed in Sect.\,\ref{s-abundances}, the O, Si, Cr, and Fe abundances
that we derive from the COS spectroscopy are lower than those
determined by \citet{vennesetal11-1} and
\citet{melisetal11-1}. However, the discussion of the nature of the
planetary material is usually based on relative metal-to-metal
abundance ratios \citep{nittleretal04-1}, which are more robust than 
absolute abundances measurements. Figure\,\ref{f-mm} compares the
metal abundances determined for GALEX\,1931+0117 normalised with respect
to Si, and relative to the corresponding ratios for the chemical
composition of the bulk Earth. It is evident that our metal-to-Si
ratios are consistent with those of \citet{melisetal11-1}, whereas
the Mg/Si, Fe/Si, and Ca/Si ratios of \citet{vennesetal11-1} are
systematically lower. 

\section{The nature and origin of the circumstellar material}
\label{s-debrisnature}
The four white dwarfs studied here have diffusion time scales of a few
days (Sect.\,\ref{s-diffusion}), and we can therefore safely assume
that we observe them in accretion-diffusion equilibrium. In other
words, the abundances of the circumstellar debris can be determined
from the photospheric analysis without any additional assumptions
regarding the history of the accretion rate that are necessary for
stars with very long diffusion time scales
\citep[e.g.][]{kleinetal11-1}. In what follows, we discuss the
abundances of the circumstellar debris normalised to Si, the main
rock-forming element, as is common use for solar-system objects
\citep[e.g.][]{lodders+fegley11-1}. 

Figure\,\ref{f-mm} (right panel) illustrates the metal-to-Si 
ratios of the planetary debris around the four white dwarfs
relative to the same abundances of the bulk Earth model by
\citet{mcdonough00-1}. The first striking observation is that the C/Si
ratios of all four stars (including one upper limit) are much lower
than that of CI chondrites, and in fact agree within their errors
with the C/Si value of the bulk Earth model. While the C abundance of
the bulk Earth is subject to some model-dependent assumptions (see the
left panel of Fig.\,\ref{f-mm} for an alternative chemical model of
the Earth by \citealt{allegreetal01-1}), these uncertainties are
comparable to the errors in our abundance determinations. 

For comparison, we include in Fig.\,\ref{f-mm} the abundance ratios of
three white dwarfs that accrete from the wind of a close M-dwarf
companion, that were also observed as part of our COS snapshot
programme\footnote{A more detailed discussion of these binaries will
  be published elsewhere. Here, they merely serve as ``abundance
  standard white dwarfs'' which accrete material with abundance ratios
  that are expected to be close to solar, i.e. rich in
  volatiles.}. The only elements detected in the COS spectra of these
three stars are C, O, Si and S, and they exhibit high abundances in C
and S, as expected for the accretion of solar-like material.  The
extremely low abundances of the volatile C found for the debris around
the four white dwarfs strongly underlines its rocky nature. This
corroborates the previous studies of \citet{jura06-1} and
\citet{juraetal12-1}, who found strong evidence for substantial
depletion of C around three DB white dwarfs.

However, Fig.\,\ref{f-mm} also shows that there is a significant
scatter among the individual abundances for a given element. Among the
four targets, the abundances of the debris in SDSS\,1228+1040 most
closely resembles those of the bulk Earth. PG\,1015+161 stands out by
having all detected elements over-abundant with respect to Si, when
compared to the bulk Earth. An interesting trend is seen in
PG\,0843+516, where Fe, Ni, and S are significantly over-abundant,
and, in fact, broadly consistent with the abundance ratios of the core
Earth model. In particular, the volatile S is extremely overabundant
with respect to C, compared to the bulk silicate Earth. In melts, S
will form FeS, and hence be depleted from remaining minerals. The
affinity of S to Fe is thought to be the reason for the depletion of S
in the silicate mantle of the Earth, as it will have settled into the
Earth's core in the form of iron sulfide \citep{ahrens79-1,
  dreibus+palme96-1}. Similarly, also Cr is significantly
over-abundant in PG\,0843+516 with respect to the bulk Earth. While Cr
is a moderately volatile element, the depletion of Cr in the silicate
Earth is thought to be due to partitioning into the Earth's core
\citep{moynieretal11-2}. Finally, the refractory lithophile Al is
under-abundant compared to the silicate Earth. Thus, the abundance
pattern seen in PG\,0843+516 suggests that the planetary debris is
rich in material that has undergone at least partial melting, and
possibly differentiation. A possible test of this hypothesis would be
a measurement of the abundance of Zn, a lithophile element with a
similar volatility as S that is not depleted into iron melt
\citep{lodders03-1}, and it will be important to test whether the
refractory lithophile Ca is depleted at a similar level as Al. The
most promising feature to measure the Zn abundances is the
\Lines{Zn}{ii}{2026,2062} resonance doublet, and \Ion{Ca}{ii}\,K
should be easily detectable in high-resolution optical spectroscopy.

\begin{figure*}
\centerline{\includegraphics[width=1.5\columnwidth]{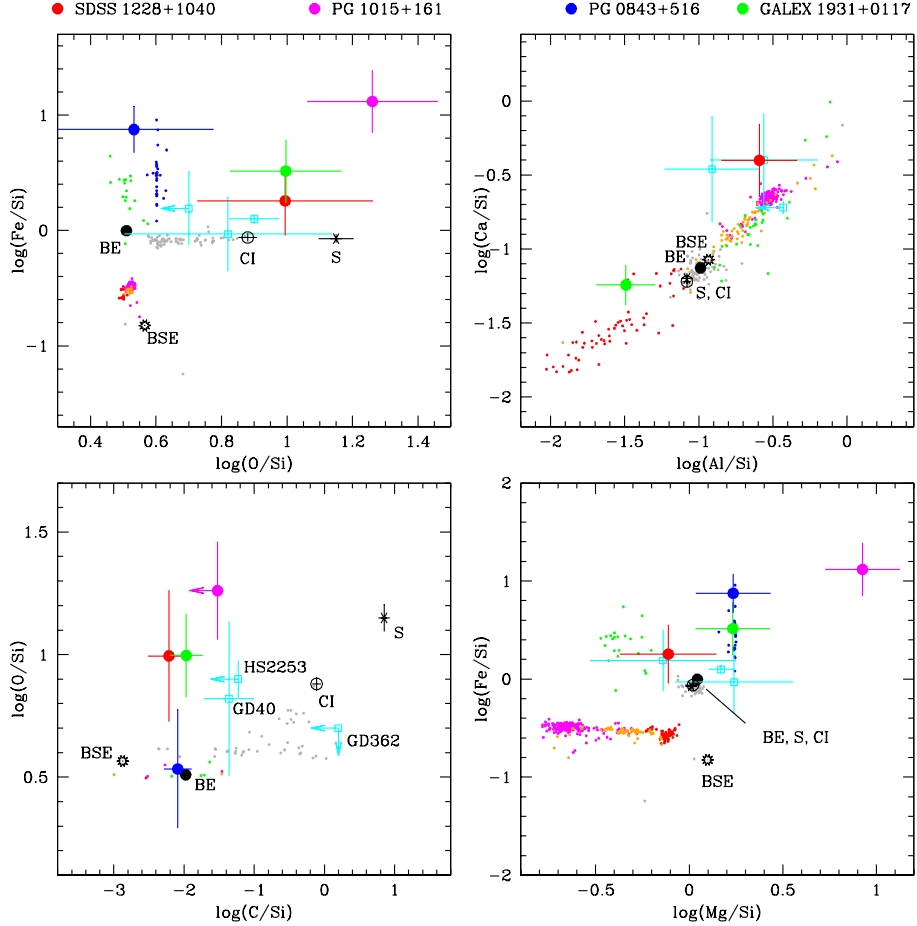}}
\caption{\label{f-mm_mm} The chemical abundances of planetary debris,
  determined from the photospheric studies of polluted white dwarfs,
  reveal a large degree of diversity. The four panels illustrate a
  range of metal-to-Si number abundance ratios for the four stars
  analysed in this paper, compared to those of bulk Earth and bulk
  silicate Earth (BE and BSE; \citealt{mcdonough00-1}), solar
  abundances and CI chondrites (S and CI; \citealt{lodders03-1}),
  and several meteorite classes (gray\,=\,carbonaceous Chondrites,
  green\,=\,Mesoderites, blue\,=\,Pallasites, red\,=\,Diogenites,
  orange\,=\,Howardites, magenta\,=\,Eucrites;
  \citealt{nittleretal04-1}). Also shown, in light blue, are the
  abundance ratios for the polluted DB white dwarfs GD\,362
  \citep{zuckermanetal07-1}, GD\,40 \citep{juraetal12-1}, and
  HS\,2253+8023 \citep{kleinetal11-1}. }
\end{figure*}

To further explore the chemical diversity of the planetary debris
around the four white dwarfs studied here, we compare pairwise a range
of metal-to-Si abundance ratios with those of the bulk Earth and bulk
silicate Earth \citep{mcdonough00-1}, as well as with those of a
variety of meteorites (taken from \citealt{nittleretal04-1}). We
inspect first the relative abundances of Al and Ca, which are two of
the three most abundant refractory lithophile elements (the third one
being Ti), i.e. elements that sublimate only at very high
temperatures, and that do not enter the core in the case of
differentiation. Therefore, the Al/Ca ratio is nearly constant across
most classes of meteorites, and hence, the Al/Si values determined from
many solar-system bodies follows a linear correlation with Ca/Si
(Fig.\,\ref{f-mm_mm}, top right). Finding that the abundances for the
debris discs, where Al, Ca, and Si are available, generally follow
that trend is reassuring, as large variations in the relative Al and
Ca abundances would cast doubts on the overall methodology using white
dwarf photospheres as proxies for the abundances of the circumstellar
material.

The relative abundances of O, Si, Mg, and Fe, which are the major
constituents of the terrestrial planets in the solar system, show
substantial variations between different meteorite groups
(Fig.\,\ref{f-mm_mm}, top left and bottom right panels), and at least
as much scatter between the individual white dwarfs. The difficulty
with these elements is that they form a range of different minerals
(metal oxides), depending on the prevailing pressure and
temperature. Iron in particular may occur as pure metal, alloy, or
mineral, and is subject to differentiation into planetary
cores. Oxygen, on the other hand, can be be locked in a wide range of
oxides (see the discussion by \citealt{kleinetal10-1}), or potentially
water \citep{kleinetal10-1, jura+xu10-1, farihietal11-1,
  jura+xu12-1}. Therefore, the relative abundances of O, Si, Mg, and
Fe will vary according to the processing that material underwent
(e.g. condensation, melting, and differentiation), and it is maybe not
too surprising to find that the debris around white dwarfs exhibits at
substantial degree of diversity, as it represents different planetary
systems formed around different stars. We note that the debris at
PG\,0843+516 falls close to the abundance ratios of Pallasites, a
class of stony-iron meteorites. This further supports our hypothesis
that PG\,0843+516 is accreting material in which iron has undergone
(partial) melting. 

Another interesting pair of elements is C and O (Fig.\,\ref{f-mm_mm},
lower left panel). The possible range of the C/O ratio among
exo-planets has been subject to intense discussion. 
It is thought that for $\mathrm{C/O}>0.8$ in the proto-planetary
discs, the ambient chemistry will favour solid ``carbon planets'',
that are dominated by carbides rather than oxides
\citep{kuchner+seager05-1}. The possible existence of carbon planets
has gained some support by the recent report of a C/O value exceeding
unity in the atmosphere of the transiting hot Jupiter WASP-12b
\citep{madhusudhanetal11-1}, and by abundance studies that found a
significant fraction of exo-planet host stars having
$\mathrm{C/O}>0.8$ (\citealt{petigura+marcy11-1, delgademenaetal10-1}),
but see \citet{fortney12-1} for a critical discussion.

Planetary debris at white dwarfs provides a unique opportunity to
probe the C/O ratio of exo-terrestrial material. However, measuring C
abundances in white dwarfs is challenging, as the optical detection of
carbon in cool white dwarfs is usually related to dredge-up from the
core rather than external pollution \citep[e.g.][]{dufouretal05-1,
  koester+knist06-1, desharnaisetal08-1}. At higher temperatures,
where convective dredge-up can be excluded, suitable lines of C are
only found at ultraviolet wavelengths.  As mentioned above, the four
stars studied here have very similar (low) C/Si ratios, but do show a
range of O/Si ratios. Nevertheless, the debris around all four stars
studies here, as well as GD\,40 \citep{juraetal12-1}, have
$-3\la\log(\mathrm{C/O})\la-2.3$, very similar to the bulk silicate
Earth, $\log(\mathrm{C/O})\simeq-2.5$, and are hence representative of
solar system minerals.

\section{Accretion rates}
\label{s-mdot}

Estimating accretion rates for metal-polluted white dwarfs is
notoriously difficult, as it is based on scaling from the elements
detected in the photosphere to an assumed bulk composition of the
accreted material. In addition, in the case of white dwarfs with
significant convective envelope masses, only the average accretion
rate over the diffusion time scale can be obtained. 

\citet{koester+wilken06-1} calculated accretion rates for 38 DAZ white
dwarfs based on the abundance of Ca, and adopting solar abundances for
the accreting material. For PG\,1015+161, these assumptions implied
$\dot M\simeq2\times10^{11}\,\mathrm{g\,s^{-1}}$. Since then, it has become
increasingly clear that many, if not most, metal-polluted (single)
white dwarfs accrete volatile-depleted material from circumstellar
planetary debris.  \citet{farihietal09-1} estimated accretion rates
for 53 metal-polluted white dwarfs following the prescription of
\citet{koester+wilken06-1}, but scaling the results by the typical
gas-to-dust ratio in the interstellar medium to account for the
absence of H and He in the accreted debris, resulting in $\dot
M\simeq2\times10^{9}\,\mathrm{g\,s^{-1}}$ for PG\,1015+161.

The uncertainty in the estimated accretion rates can be greatly
reduced if photospheric abundances for the major constituents of the
debris material can be measured. While we do not detect all elements
that are likely present in the circumstellar debris at the four white
dwarfs studied here, we have determined the accretion rates of all the
major elements, in particular O, Si, Mg, and Fe
(Sect.\,\ref{s-diffusion}). The accretion rates of all detected
elements, as well as their sum are given in Table\,~\ref{t-fluxes},
and are illustrated in Fig.~\ref{f-mdot}. For PG\,1015+161, we find
$\dot M\simeq1.7\times10^8\,\mathrm{g\,s^{-1}}$, which is strictly
speaking a lower limit, however, the undetected elements (e.g. Al, S,
Ti, Mn, Cr) are unlikely to contribute more than 10\% of the total
accretion rate. Similarly, we find the accretion rates of
PG\,0843+516, SDSS\,1228+1040, and GALEX\,1931+0117 to be $\dot
M\simeq1.0\times10^9\,\mathrm{g\,s^{-1}}$,
$5.6\times10^8\,\mathrm{g\,s^{-1}}$, and
$1.5\times10^9\,\mathrm{g\,s^{-1}}$, respectively.

\section{Hot circumstellar gas}
\label{s-hotgas}
The discs around white dwarfs are passive, i.e. their emission is
solely due to the thermal reprocessing of intercepted stellar
flux. The inner disc radius where typical dust grains will rapidly
sublimate is determined by the luminosity of the white dwarf
\citep{vonhippeletal07-1}. The gaseous material will viscously spread,
both flowing inwards onto the white dwarf, and outwards over the dusty
disc, potentially accelerating the inwards migration of the dust via
aerodynamic drag \citep{rafikov11-2}.  While gaseous material orbiting
at radii coincident with circumstellar dust is observed in a number of
systems in the form of double-peaked emission lines
\citep{gaensickeetal06-3, gaensickeetal07-1, gaensickeetal08-1,
  brinkworthetal09-1, brinkworthetal-12, melisetal11-1, melisetal12-1,
  farihietal12-1, dufouretal12-1}, there has yet been no detection of
gaseous material well inside the sublimation radius.

Inspection of Fig.\,~\ref{f-fit1} reveals that the strength of the
\Lines{Si}{iv}{1394,1403} doublet in PG\,0843+516 is extremely
under-predicted by the photospheric model. These \Ion{Si}{iv} lines
correspond to the highest ionisation energy of all transitions
detected in the COS spectrum.  For the
temperature and the Si abundance of PG\,0843+516, the observed
strength of the \Ion{Si}{iv} lines is absolutely incompatible with a
purely photospheric origin.  The most plausible explanation is that
there is additional absorption along the line of sight, associated
with hot gas close to the white dwarf that is optically thin except
for the strong resonance lines of high-ionisation species, such as
\Ion{Si}{iv}. In fact, extremely similar features were found in the
far-ultraviolet observations of cataclysmic variables, i.e.  white
dwarfs that accrete from a (hydrogen-rich) accretion disc that is in
turn fed by Roche-lobe overflow of a close M-dwarf companion.
\textit{HST}/GHRS and \textit{FUSE} spectroscopy of the white dwarf in
U\,Gem contains very strong absorption of \Lines{N}{v}{1239,1243} and
\Lines{O}{vi}{1032,1038} that can not form in the $\simeq30\,000$\,K
photosphere, as well as excess absorption in \Lines{Si}{II}{1394,1403}
\citep{sionetal98-1, long+gilliland99-1, longetal06-1}. All three
high-ionisation doublets are red-shifted with respect to the systemic
velocity of the white dwarf, but somewhat less so than the
lower-ionisation photospheric lines, which are subject to the
gravitational redshift at the photospheric radius. These observations
were interpreted as evidence for a hot ($\sim80\,000$\,K) layer of gas
sufficiently close to the white dwarf to still experience a noticeable
gravitational redshift.
Measuring the central wavelengths of the strong \Lines{Si}{iv}{1394,
  1403} lines in PG\,0843+516, we find that they are blue-shifted with
respect to the photospheric features by $\simeq25$\,\kms, which
implies a height of $\simeq1.5$ white dwarf radii above the white dwarf
surface. This assumes that there is no significant flow velocity,
which seems reasonably well justified given the symmetric shape of the
\Ion{Si}{iv} profiles.

A discrepancy between the best-fit white dwarf model and the region
around the \Ion{Si}{iv} doublet is also seen in the COS spectrum of
SDSS\,1228+1040 (Fig.~\ref{f-fit2}, bottom left panel), however, in
this star, the additional absorption is rather weak. These additional
absorption features are clearly blue-shifted with respect to the
photospheric lines, however, the relatively low signal-to-noise ratio
of the spectrum prevents an accurate determination of this offset.

For PG\,1015+161 and GALEX\,1931+0117, the photospheric fits match the
observed \Ion{Si}{iv} lines well, i.e. there is no evidence for any
additional absorption component.  Given that these two stars have,
respectively, the lowest and highest accretion rate of our small
sample (Sect.\,\ref{s-mdot}), there seems to be no clear correlation
between the detection of absorption from highly ionised gas to the
mass flow rate onto the white dwarf.  A key difference between the two
stars where circumstellar \Ion{Si}{iv} absorption is detected is that
SDSS\,1228+1040 also shows strong \textit{emission} lines from
circumstellar gas, which indicate a relatively high inclination of the
accretion disc. In contrast, no gaseous emission is found in
PG\,0843+516 (G\"ansicke et al. in prep).  Identifying additional
absorption features from these hot layers of gas would provide
substantial constraints on the physical parameters in the
corresponding regions. The strongest line seen in cataclysmic
variables, \Ion{N}{v}, is naturally absent in the white dwarfs
accreting rocky debris\footnote{For completeness, we note that
  circumstellar high-ionisation absorption lines have also been found
  around a number of hot white dwarfs \citep{bannisteretal03-1,
    dickinsonetal12-1}. However, the origin of the circumstellar
  material is not clear, and the detection of strong C lines suggests
  a different nature compared to the rocky debris found around the
  stars studied here.}, but the \Lines{O}{vi}{1032,1038} doublet
detected in U\,Gem \citep{longetal06-1} is a promising candidate.

\section{Conclusions}

Recent years have seen a surge of interest in the evolution of
extra-solar planetary systems through the late phases in the lifes of
their host stars \citep[e.g.][]{burleighetal02-1, debesetal02-1,
  villaver+livio07-1, villaver+livo09-1, nordhausetal10-1,
  distefanoetal10-1}. While no planet has yet been discovered orbiting
a white dwarf \citep{hoganetal09-1, faedietal11-1}, significant
progress has been made in the discovery and understanding of planetary
debris discs around white dwarfs.

Our COS study substantially increases the number of polluted white
dwarfs for which a wide range of chemical elements have been
detected. We find that the C/Si ratio is consistent with that of the
bulk Earth, which confirms the rocky nature of the debris at these
white dwarfs, and their C/O values are typical of minerals dominated by Fe
and Mg silicates. There is so far no detection of planetary debris at
white dwarfs that has a large C/O ratio which would be indicative of
silicon carbide-based minerals.  The abundances of planetary material
found around white dwarfs show a large diversity, comparable to, or
exceeding that seen among different meteorite classes in the solar
system. We find that the Al/Ca ratio follows a similar trend as
observed among solar system objects, which suggests that processing of
proto- and post-planetary material follows similar underlying
principles. A particularly interesting pattern is found in
PG\,0843+516, where over-abundances of S, Cr, Fe, and Ni are
suggestive of the accretion of material that underwent melting and
possibly differentiation. Extending the abundance studies of
metal-polluted white dwarfs both in detail and number will provide
further insight into the diversity of exo-terrestrial material, and
guide the understanding of terrestrial exo-planet formation
\citep{bondetal10-1, carter-bondetal12-1}.

\vspace*{-3ex}
\section*{Acknowledgements}
We gratefully acknowledge Larry Nittler for sharing his meteorite
abundance data with us, and William Januszewski, Charles Proffitt, and
Elena Mason for their tireless efforts in the implementation of the
\textit{HST} program. 
D.K. wants to thank P.-E. Tremblay and P. Bergeron for sharing their
new calculations of the hydrogen Lyman and Balmer line Stark profiles.
Based on observations made with the NASA/ESA Hubble Space Telescope,
obtained at the Space Telescope Science Institute, which is operated
by the Association of Universities for Research in Astronomy, Inc.,
under NASA contract NAS 5-26555. These observations are associated
with program \#11561, \#12169 and \#12474. Also based on observations
made with ESO Telescopes at the La Silla Paranal Observatory under
programme ID 79.C-0085, 81.C-0466, 82.C-0495, 383.C-0695.
We thank the anonymous referee for a constructive report. 

\bibliographystyle{mn_new}

\bsp

\label{lastpage}

\end{document}